\documentclass[a4paper,twoside,12pt]{article}
\input{obsjkt.sty}

\newcommand{\Msun}{\ensuremath{~{\rm M}_\odot}}                   
\newcommand{\Rsun}{\ensuremath{~{\rm R}_\odot}}                   
\newcommand{\rhosun}{\ensuremath{~\rho_\odot}}                    
\newcommand{\Teff}{\ensuremath{T_{\rm eff}}}                      
\newcommand{\TeffA}{\ensuremath{T_{\rm eff,A}}}                   
\newcommand{\TeffB}{\ensuremath{T_{\rm eff,B}}}                   
\newcommand{\EBV}{\ensuremath{E(B\!-\!V)}}                        
\newcommand{\Grp}{\ensuremath{G_{\rm RP}}}                        
\newcommand{\degr}{\ensuremath{^\circ}}                           
\renewcommand{\kms}{~km~s$^{-1}$}                                 
\newcommand{\am}{\ensuremath{^\prime}}                            
\newcommand{\as}{\ensuremath{^{\prime\prime}}}                    
\newcommand{\etal}{\textit{et al.}}                               
\newcommand{\kepler}{\textit{Kepler}}                             
\newcommand{\gaia}{\textit{Gaia}}                                 
\newcommand{\targ}{CW~CMa}
\newcommand{\targfull}{CW Canis Majoris}

\newcommand{\Msunnom}{\hbox{$\mathcal{M}^{\rm N}_\odot$}}
\newcommand{\Rsunnom}{\hbox{$\mathcal{R}^{\rm N}_\odot$}}
\newcommand{\Lsunnom}{\hbox{$\mathcal{L}^{\rm N}_\odot$}}

\usepackage{rotating}

\begin{document} 

\OBSheader{Rediscussion of eclipsing binaries: \targ}{J.\ Southworth}{2026 October}

\OBStitle{Rediscussion of eclipsing binaries. Paper 33. \\ The A-type system CW Canis Majoris in the PLATO southern field}

\OBSauth{John Southworth and Stephen Overall}

\OBSinstone{Astrophysics Group, Keele University, Staffordshire, ST5 5BG, UK}


\OBSabstract{\targ\ is a detached eclipsing binary containing two A0~V stars in a circular orbit of period 2.12~d. We analyse the light-curves of this system from the Transiting Exoplanet Survey Satellite (TESS) together with published radial velocity measurements to determine the masses and radii of the component stars to high precision, obtaining $M_{\rm A} = 2.087 \pm 0.022$\Msun, $M_{\rm B} = 1.972 \pm 0.024$\Msun, $R_{\rm A} = 1.841 \pm 0.012$\Rsun\ and $R_{\rm B} = 1.748 \pm 0.014$\Rsun. We detect a changing orbital period over the four available TESS sectors indicative of the presence of a companion on a wider orbit. We develop a new method to determine the effective temperatures of the component stars using their measured radii, the \gaia\ parallax, and surface-brightness calibrations to reduce the degeneracy between temperature and extinction. The method gives temperatures of $9850 \pm 280$~K and $9570 \pm 270$~K and an interstellar extinction of $A_V = 0.32 \pm 0.06$~mag. The properties of the system can be matched by theoretical models for a subsolar metal abundance ($Z=0.010$) and an age of $400 \pm 20$~Myr. \targ\ has been selected as a target for the PLATO satellite, with which it will be observed for at least two years.}


\section*{Introduction}

Detached eclipsing binary star systems (DEBs) are of fundamental importance to stellar astrophysics as the primary source of directly measured properties of stars that have evolved as if they were single. Measurement of their physical properties traditionally relied on obtaining extensive observations -- light-curves from small telescopes and high-resolution spectroscopy from typically larger telescopes -- for each DEB, the large time and effort required limiting the number of objects that could be studied. At the time of writing, a total of 374 DEBs have precise mass and radius measurements for their component stars and have been entered into the Detached Eclipsing Binary Catalogue \cite{Me15aspc} (DEBCat\footnote{\texttt{https://www.astro.keele.ac.uk/jkt/debcat/}}). Whilst this sounds like a good number, it should be remembered that the first determination of mass and radius for a DEB, for $\beta$~Aurigae \cite{Stebbins11apj}, occurred as long ago as 1911.

The availability of extensive photometry from a series of space-based survey telescopes (CoRoT, \kepler, K2, TESS, and next PLATO) has changed this picture, making it no longer necessary to perform long observing runs to obtain the light-curves needed to analyse a DEB \cite{Me21univ}. This has prompted the current series of papers \cite{Me20obs} in which primarily space-based light-curves are used to refine the properties of known DEBs. In the current work we study \targ\ (Table~\ref{tab:info}), whose properties were established in 1982 based on light-curves much inferior to those now available.


\section*{\targfull}


The discovery of light variations in \targ\ was announced by Van Hoof \cite{Vanhoof49plagl}, from blinking photographic plates taken by Van Gent in Johannesburg. Van Hoof specified an orbital period of 1.0589826~d and ascribed its discovery to Deurinck. This period turns out to be half the true value, a common error when a limited amount of data is available for a DEB with a circular orbit and primary and secondary eclipses of similar depth.

A complete light-curve was established by Williamon \cite{Williamon76aj}, who obtained 692 photoelectric observations in each of the $UBV$ bands. Williamon corrected the orbital period to 2.11797737~d and analysed the data using the method of Russell \& Merrill \cite{RussellMerrill52book}. These data were reanalysed by Mezzetti \etal\ \cite{Mezzetti+80aas} using the {\sc wink} code \cite{Wood71aj}.

Lacy \cite{Lacy82apj} (hereafter L82) obtained 20 spectra using the McDonald 2.7-m telescope, coud\'e spectrograph, and a Reticon or Digicon detector. He combined the radial velocities (RVs) measured from these spectra with the photometric results from Williamon to determine the masses of the stars to 0.8\% and the radii of the stars to 2\% (primary component, hereafter star~A) and 4\% (secondary component, hereafter star~B) precision. The rotational velocities of the two stars were measured as $46 \pm 3$\kms\ and $45 \pm 3$\kms, respectively. No detailed analysis of \targ\ has been presented since.

\begin{table}[t]
\caption{\em Basic information on \targfull. 
The $BV$ magnitudes are each the mean of 154 individual measurements \cite{Hog+00aa} distributed approximately 
randomly in orbital phase. The $JHK_s$ magnitudes from 2MASS \cite{Cutri+03book} were obtained at an orbital 
phase of 0.41, which is close to but confidently before the beginning of primary eclipse. \label{tab:info}}
\centering
\begin{tabular}{lll}
{\em Property}                            & {\em Value}                 & {\em Reference}                      \\[3pt]
Right ascension (J2000)                   & 07 21 52.53                 & \citenum{Gaia23aa}                   \\
Declination (J2000)                       & $-$23 47 37.4               & \citenum{Gaia23aa}                   \\
Henry Draper designation                  & HD 57802                    & \citenum{CannonPickering19anhar}     \\
\textit{Tycho} designation                & TYC 6537-1289-1             & \citenum{Hog+00aa}                   \\
\textit{Gaia} DR3 designation             & 5617822243996018944         & \citenum{Gaia21aa}                   \\
\textit{Gaia} DR3 parallax (mas)          & $2.9739 \pm 0.0199$         & \citenum{Gaia21aa}                   \\          
TESS\ Input Catalog designation           & TIC 412254866               & \citenum{Stassun+19aj}               \\
$B$ magnitude                             & $8.680 \pm 0.010$           & \citenum{Hog+00aa}                   \\          
$V$ magnitude                             & $8.602 \pm 0.012$           & \citenum{Hog+00aa}                   \\          
$J$ magnitude                             & $8.370 \pm 0.026$           & \citenum{Cutri+03book}               \\
$H$ magnitude                             & $8.380 \pm 0.042$           & \citenum{Cutri+03book}               \\
$K_s$ magnitude                           & $8.378 \pm 0.023$           & \citenum{Cutri+03book}               \\
Spectral type                             & A0~V                        & \citenum{HoukSmithmoore88book}       \\[3pt]       
\end{tabular}
\end{table}



\begin{figure}[t] \centering \includegraphics[width=\textwidth]{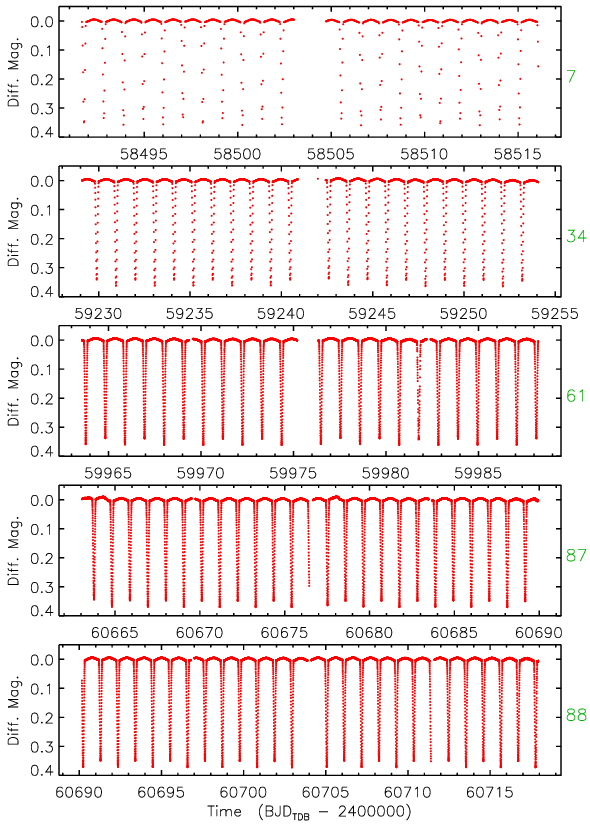} \\
\caption{\label{fig:time} Light-curves of \targ\ from all five TESS sectors observed 
so far. The flux measurements have been converted to magnitude units and the median 
subtracted. The sector numbers are given to the right of each panel.} \end{figure}


\section*{Photometric observations}


\targ\ has been observed by the Transiting Exoplanet Survey Satellite (TESS \cite{Ricker+15jatis}) mission in five sectors so far. Observations in a further four sectors (108--111) are expected in the near future (September to December 2026). The observations in sector 7 were summed to a cadence of 1800~s before transmission to Earth, those in sector 34 were summed to 600~s cadence, and the remainder were summed to 200~s cadence. The available data products also vary: the data from sectors 7, 34 and 61 have been reduced by the TESS Science Processing Center (SPOC \cite{Jenkins+16spie}), and those from sectors 87 and 88 using the \textsc{Quick Look Pipeline} (QLP \cite{Huang+20rnaas}).

The data from each of the five sectors were downloaded from the Mikulski Archive for Space Telescopes (MAST\footnote{\texttt{https://mast.stsci.edu/portal/Mashup/Clients/Mast/Portal.html}}) using the {\sc lightkurve} package \cite{Lightkurve18}, with the ``hard'' quality flag specified to reject data of lower quality. The median magnitude was then subtracted for each sector to normalise the data to zero outside eclipse. The light-curves are shown in Fig.~\ref{fig:time}.

\targ\ is close to the galactic plane, in a field of relatively high stellar density. A query of the \gaia\ DR3 database\footnote{\texttt{https://vizier.cds.unistra.fr/viz-bin/VizieR-3?-source=I/355/gaiadr3}} for sources within 2~arcmin returned 248 matches in addition to \targ\ itself, the brightest of which is 4.50~mag fainter in the \Grp\ band. However, there are bright stars at angular separations of 2.1\am\ and 3.9\am, which are fainter by only 1.09 and 0.77~mag. We therefore have reason to suspect that contamination of the TESS light-curve by nearby stars is not negligible


\section*{Light-curve analysis}

Our analysis concentrated on the data from TESS sector 61. This is because the light-curves from sectors 7 and 34 have poorer time resolution, and those from sectors 87 and 88 are from the QLP pipeline so have unreliable eclipse depths \cite{Me26obs3}. 

\begin{figure}[t] \centering \includegraphics[width=\textwidth]{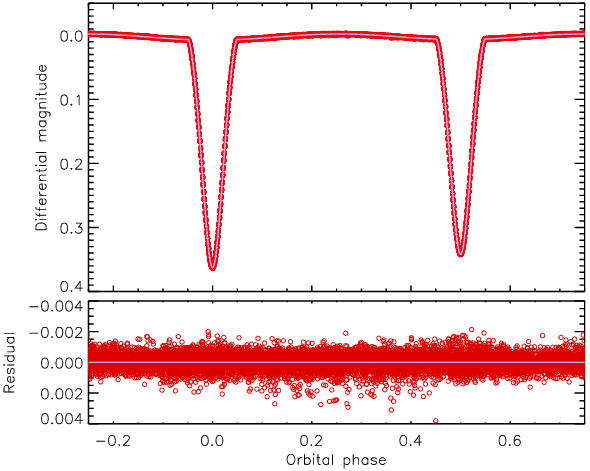} \\
\caption{\label{fig:phase} {\sc jktebop} best fit to the light-curve of \targ\ from TESS 
sector 61 (upper panel) and the residuals of the fit (lower panel). The data are shown as 
open red circles. The best fit is shown using a light blue solid line.} \end{figure}

We modelled the sector 61 light-curve using version 45 of the {\sc jktebop}\footnote{\texttt{http://www.astro.keele.ac.uk/jkt/codes/jktebop.html}} code \cite{Me++04mn2,Me13aa}, fitting for the orbital period ($P$), a reference time of primary minimum ($T_0$), the sum and ratio of the fractional radii ($r_{\rm A}+r_{\rm B}$ and $k = r_{\rm B}/r_{\rm A}$), the central surface brightness ratio ($J$), third light ($L_3$), and orbital inclination ($i$). A circular orbit was assumed due to the short orbital period, the appearance of the light-curve, and the fact that it gave an excellent fit to the data. 
We included limb darkening (LD) using the power-2 law \cite{Hestroffer97aa,Maxted18aa,Me23obs2} with coefficients taken from theoretical predictions \cite{ClaretSouthworth22aa,ClaretSouthworth23aa}. We fitted for the $c$ coefficients for both stars, and left the $\alpha$ coefficients fixed at the theoretical value. We also applied four first-order polynomials, one for each quarter of the sector, to the brightness of the system and fitted for the two coefficients of each. \targ\ is a well-behaved system and we immediately obtained a good fit to the data. This is plotted in Fig.~\ref{fig:phase} and the parameters are given in Table~\ref{tab:jktebop}.

\clearpage

Uncertainties in the fitted parameters were deduced in three ways. First, we ran 1000 Monte Carlo simulations using Task 8 in {\sc jktebop} \cite{Me08mn}. Second, we ran the same number of residual-permutation simulations using Task 9 in {\sc jktebop} \cite{Me08mn}. Third, we subdivided the light-curve into seven sections, each containing three contiguous eclipses and the out-of-eclipse data between and around them. No data were allocated to multiple sections, and one eclipse was ignored as it was surplus to requirements. Each section was then fitted in the same way as before, except that the orbital period was fixed to the best-fitting value found above. The uncertainty of each parameter was taken to be the standard deviation of the individual values divided by $\sqrt{7}$. 

The three sets of uncertainties were found to be in very good agreement. The subdivision uncertainties were the highest for $k$, $J$, $r_{\rm A}$, $r_{\rm B}$ and the light ratio $\ell_{\rm B}/\ell_{\rm A}$; the Monte-Carlo uncertainties were the highest for the remaining parameters. The ratio of the radii is comparatively poorly-determined, as is typical for DEBs with deep but partial eclipses, but it is sufficiently well-measured that a spectroscopic light ratio is unlikely to significantly improve the situation. The highest uncertainty for each parameter is reported in Table~\ref{tab:jktebop}.

\begin{table} \centering
\caption{\em \label{tab:jktebop} Photometric parameters of \targ\ measured using 
{\sc jktebop} from the TESS light-curves. The error bars are 1$\sigma$ and are
the largest of three options calculated for each parameter.}
\begin{tabular}{l r@{\,$\pm$\,}l r@{\,$\pm$\,}l}
{\em Parameter}                               & \multicolumn{2}{c}{\em~~~~Value} \\[3pt]
{\it Fitted parameters:} \\
Orbital inclination (\degr)                   &      83.279    &   0.017         \\
Sum of the fractional radii                   &       0.32327  &   0.00019       \\
Ratio of the radii                            &       0.9744   &   0.0012        \\
Central surface brightness ratio              &       0.9538   &   0.0019        \\
Third light                                   &       0.0294   &   0.0011        \\
LD coefficient $c$ of star~A                  &       0.555    &   0.015         \\
LD coefficient $c$ of star~B                  &       0.578    &   0.015         \\
LD coefficient $\alpha$ of star~A  & \multicolumn{2}{c}{~~~~~~~~~0.4490 (fixed)} \\
LD coefficient $\alpha$ of star~B  & \multicolumn{2}{c}{~~~~~~~~~0.4530 (fixed)} \\
Midpoint of primary eclipse (BJD$_{\rm TDB}$) & 2459976.475600 &   0.000005      \\
{\it Derived parameters:} \\
Fractional radius of star~A                   &       0.1663   &   0.0009        \\
Fractional radius of star~B                   &       0.1570   &   0.0011        \\ 
Light ratio $\ell_{\rm B}/\ell_{\rm A}$       &       0.847    &   0.022         \\
\end{tabular}
\end{table}


\section*{Orbital ephemeris}

\begin{table} \centering
\caption{\em Times of mid-eclipse for \targ\ and their residuals versus the linear ephemeris. \label{tab:tmin}}
\setlength{\tabcolsep}{10pt}
\begin{tabular}{rrrrr}
{\em Orbital} & {\em Eclipse time}  & {\em Uncertainty} & {\em Residual} & {\em TESS}   \\
{\em cycle}   & {\em (BJD$_{TDB}$)} & {\em (d)}         & {\em (d)}      & {\em sector} \\[3pt]
$-696.0$ & 2458502.364338 & 0.000014 & $ 0.000085$ & 07  \\   
$-348.0$ & 2459239.419957 & 0.000017 & $ 0.000013$ & 34  \\   
$   0.0$ & 2459976.475600 & 0.000005 & $-0.000035$ & 61  \\   
$ 330.0$ & 2460675.407756 & 0.000009 & $ 0.000000$ & 87  \\   
$ 343.0$ & 2460702.941460 & 0.000004 & $ 0.000015$ & 88  \\   
\end{tabular}
\end{table}

\begin{figure}[t] \centering \includegraphics[width=\textwidth]{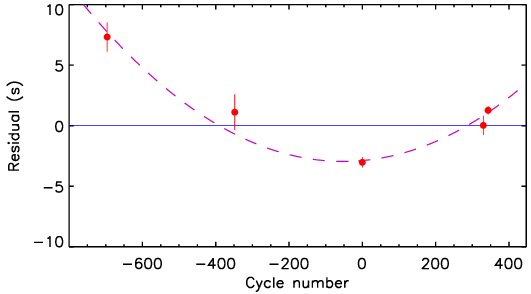} \\
\caption{\label{fig:tmin} Residuals of the times of minimum light from Table~\ref{tab:tmin} 
versus the best-fitting linear ephemeris. Red filled circles indicate the timings from the
TESS data found in the current work. The blue solid line indicates a residual of zero, 
and the lilac dashed line shows the best fit for the quadratic ephemeris.} \end{figure}

Up to this point in our analysis we had not used any times of eclipse from the literature, obtaining our information on the orbital ephemeris solely from the TESS sector 61 light-curve. We therefore fitted the other four TESS sectors using {\sc jktebop} to measure one effective time of primary eclipse per sector. These timings come from a fit to all eclipses in each sector, including secondary eclipses assumed to be at phase 0.5 as the orbit of \targ\ is circular. The times of eclipse are given in Table~\ref{tab:tmin} and plotted in Fig.~\ref{fig:tmin}.

We fitted these with a straight line versus epoch, under the assumption of a constant orbital period. We found, instead, a clear trend indicative of the light-time travel effect (LTTE) caused by a third body in a wider orbit. We therefore fitted the times of minimum with an ephemeris including a quadratic term. The \textit{rms} scatter of the data around the best fit decreased from 3.6~s for the linear ephemeris to 0.91~s for the quadratic ephemeris; the Bayesian Information Criterion (BIC \cite{Schwarz78,Liddle07mn}) decreased from 103.0 to 7.9; and the Akaike Information Criterion (AIC \cite{Akaike81}) decreased from 103.8 to 9.1. All three changes confirm the reality of this effect to high confidence. For reference, we fitted an ephemeris with a cubic model and obtained a scatter of 0.33~s and an almost identical BIC and AIC. We conclude that the data strongly favour a quadratic model over the linear alternative, but not a cubic model. The extremely low scatter in the times of minimum is consistent with those found in our recent work on UZ~Dra \cite{Me25obs6}, BS~Dra \cite{Me26obs2} and EY~Cep \cite{Me26obs4}, so is plausible.

We investigated this by adding published times of minimum light to our analysis \cite{Williamon76aj,Wolf+82ibvs,Kreiner++01book,Pasche17oejv,Samolyk19jaavso}. The times of minimum have much larger uncertainties than the TESS measurements, but we can conclude they are inconsistent with the quadratic ephemeris found above. A plausible explanation is that there is a third body with an orbital period significantly longer than the temporal coverage of the TESS data (6.0~yr) but significantly shorter than the time interval covered by the literature data (50.9~yr). Constraints on its period and mass could be obtained by running a grid of simulations to find those rejected by the available timings, but we prefer the simpler option of waiting until additional data become available in the near future (see below).

For convenience, we calculated a linear ephemeris based on all eclipse timings available:
\begin{equation}
\mbox{Min~I} = {\rm BJD}_{\rm TDB}~ 2459976.475635 (8) + 2.117976124 (25) E
\end{equation}
where $E$ is the number of cycles since the reference time of minimum and the bracketed quantities indicate the uncertainty in the final digit of the previous number. The times of minimum and residuals of the fit are reported in Table~\ref{tab:tmin}. This does not account for the clear quadratic trend in the TESS data, but is useful in planning future observations or analyses.


\section*{Radial velocity analysis}

\begin{figure}[t] \centering \includegraphics[width=\textwidth]{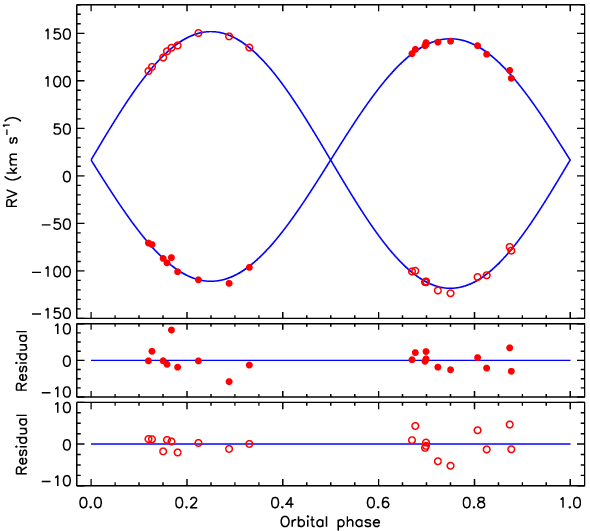} \\
\caption{\label{fig:rv} RVs of \targ\ from L82 compared to the best fit from 
{\sc jktebop} (solid blue lines). The RVs for star~A are shown with filled 
red circles, and for star~B with open red circles. The residuals are given 
in the lower panels separately for the two components.} \end{figure}

The only available RVs for \targ\ are the 20 per star published by L82, which are concentrated at the quadratures and were enough for measurement of the masses to 0.8\%. We obtained these from table~1 of L82 and performed our own analysis to confirm the measured velocity amplitudes ($K_{\rm A}$ for star~A and $K_{\rm B}$ for star~B).

\begin{sidewaystable} \centering
\caption{\em Spectroscopic orbits for \targ\ from L82 and the reanalysis of the same data in the current work.
All quantities are given in km~s$^{-1}$. \label{tab:sb}}
\setlength{\tabcolsep}{10pt}
\begin{tabular}{lccccccc}
{\em Source} & $K_{\rm A}$ & $K_{\rm B}$ & $V_{\rm \gamma}$ & $V_{\rm \gamma,A}$ & $V_{\rm \gamma,B}$ & $\sigma_{\rm A}$ & $\sigma_{\rm B}$ \\[10pt]
L82                     & $127.6  \pm 0.8 $ & $135.1  \pm 0.6 $ &                  & $16.7  \pm 0.7 $ & $16.8  \pm 0.6 $ & 3.0  & 2.5  \\
This work               & $127.60 \pm 0.70$ & $135.06 \pm 0.62$ &                  & $16.66 \pm 0.66$ & $16.74 \pm 0.57$ & 2.83 & 2.38 \\
This work (adopted)     & $127.59 \pm 0.73$ & $135.04 \pm 0.60$ & $16.71 \pm 0.43$ &                  &                  & 2.84 & 2.38 \\
\gaia\ DR3              & $131.1  \pm 1.4 $ & $137.6  \pm 1.4 $ & $19.18 \pm 0.64$ &                  &                  &      &      \\
\end{tabular}
\end{sidewaystable} 

We modelled the RVs from L82 using the linear ephemeris from the previous section, and fitting for $K_{\rm A}$, $K_{\rm B}$ and the systemic velocity ($V_{\rm \gamma}$). We allowed for a phase offset versus the ephemeris and found it to be negligible. We assumed a circular orbit based on the light-curve analysis above. A second model was identical except for allowing the two stars to have individual systemic velocities ($V_{\rm \gamma,A}$ and $V_{\rm \gamma,B}$). The uncertainties were calculated using Monte Carlo simulations \cite{Me21obs5}. The best fit for the first approach is given in Table~\ref{tab:sb} and shown in Fig.~\ref{fig:rv}. We found practically identical results for our two models and that of L82.

Spectroscopic orbits for the two stars are also available from the \gaia\ DR3 results \cite{Gaia16aa,Gaia23aa}, specifically the \texttt{tbosb2} table\footnote{\texttt{https://vizier.cds.unistra.fr/viz-bin/VizieR?-source=I/357}}. In the case of \targ\ the \texttt{tbosb2} orbits have slightly larger velocity amplitudes with error bars twice as large. The RVs on which they are based have not been published so we cannot investigate this discrepancy in detail, but note that the \texttt{tbosb2} orbits include a small but non-zero eccentricity and do not incorporate ephemeris information from the light-curve of the system. We therefore adopt our own fit of the RVs from L82, specifically that with a single $V_{\rm \gamma}$.



\section*{Effective temperature determination}

The effective temperature (\Teff) values were not determined by L82. A multitude of values has been published in the course of catalogue-construction work (e.g.\ refs.\ \citenum{Stassun+19aj,Creevey+23aa}) based on fitting a spectral energy distribution (SED) to observed apparent magnitudes, but these do not account for \targ's binarity so are unreliable. We have therefore developed our own method to obtain precise \Teff\ values.

We downloaded the apparent magnitudes of \targ\ in many passbands using the VizieR photometry tool\footnote{\texttt{https://vizier.cds.unistra.fr/vizier/sed/doc/}} developed by Anne-Camille Simon and Thomas Boch. This yielded a total of 178 magnitudes within 0.25\as\ of \targ\ after the removal of duplicates. This was further winnowed to 28 measurements by rejecting data outside the wavelength range covered by our synthetic spectra (see below), retaining only one observation per passband, and carefully pruning obvious outliers. 

The magnitudes were fitted using Markov-chain Monte Carlo (MCMC) to SEDs generated from grids\footnote{\texttt{https://svo2.cab.inta-csic.es/theory/newov2/index.php?models=bt-settl-agss}} of BtSettl-AGSS2009 synthetic spectra \cite{Allard++11aspc} to determine the \Teff\ of each star and the extinction ($A_V$) of the system. We applied Gaussian priors to the radii and surface gravities of the stars (calculated below) and the distance to the system from \gaia\ DR3 ($336.3 \pm 2.3$~pc). This resulted in measurements of $\TeffA = 10010 \pm 450$~K, $\TeffB = 9720 \pm 440$~K and $A_V = 0.32 \pm 0.10$. The error bars are quite large due to the well-known degeneracy between \Teff\ and $A_V$, which is strong in this case because the components of \targ\ are hot enough that the peak of the SED is bluer that almost all of the available apparent magnitudes.

\begin{figure}[t] \centering \includegraphics[width=\textwidth]{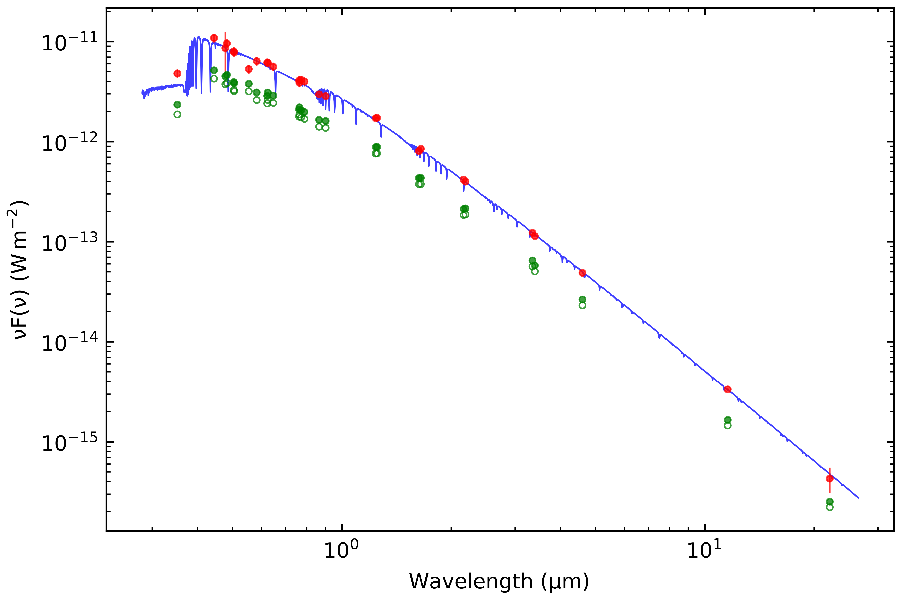} \\
\caption{\label{fig:sed} Spectral energy distribution of \targ. The observed fluxes are shown using 
red filled circles. The SED of the two stars is shown as a blue line. The green filled and open points 
give the contribution of star~A and star~B, respectively, to the fluxes in each passband.} \end{figure}

We then applied an additional restriction that the measured distance to the system must be independent of wavelength, in order to better constrain $A_V$. We determined the distance in each passband using the surface-brightness versus \Teff\ calibrations from Kervella et al.\ \cite{Kervella+04aa} and the method given by Southworth et al.\ \cite{Me++05aa}. The new results agree with the values above but are measured more precisely: $\TeffA = 9850 \pm 280$~K, $\TeffB = 9570 \pm 270$~K and $A_V = 0.320 \pm 0.059$. We adopt these values for subsequent analysis. The SED is plotted in Fig.~\ref{fig:sed}.


\section*{Physical properties of \targ}
 
We determined the physical properties of \targ\ using standard formulae and the results from our photometric and spectroscopic analyses above. The calculations were performed using the {\sc jktabsdim} code \cite{Me++05aa} and an error budget was obtained for each calculated parameter by perturbing each input parameter. The results are given in Table~\ref{tab:absdim} and show that the masses are measured to 1.2\%, and the radii to 0.8\%, precision. The synchronous rotational velocities are consistent with the values measured by L82, indicating that rotational synchronisation as well as orbital circularisation has occurred.

Our mass measurements agree well with the values from L82, as expected because our $K_{\rm A}$ and $K_{\rm B}$ are almost identical. However, our error bars for $K_{\rm A}$ and $K_{\rm B}$ are significantly larger, which suggests that the uncertainties were incompletely propagated by L82. Our radius measurements are substantially more precise, are roughly 2\% lower than those of L82, and should be preferred as they are based on much better light-curves. For reference, using the $K_{\rm A}$ and $K_{\rm B}$ from \gaia\ gives masses of $M_{\rm A} = 2.23 \pm 0.05$\Msun\ and $M_{\rm B} = 2.12 \pm 0.05$\Msun, which are approximately 2.6$\sigma$ larger than the values in Table~\ref{tab:absdim}.
     
\begin{table} \centering
\caption{\em Physical properties of \targ\ defined using the nominal solar units 
given by IAU 2015 Resolution B3 (ref.~\citenum{Prsa+16aj}). \label{tab:absdim}}
\begin{tabular}{lr@{~$\pm$~}lr@{~$\pm$~}l}
{\em Parameter}        & \multicolumn{2}{c}{\em Star A} & \multicolumn{2}{c}{\em Star B}     \\[3pt]
Mass ratio   $M_{\rm B}/M_{\rm A}$          &  \multicolumn{4}{c}{$0.9448 \pm 0.0068$}        \\
Semimajor axis of relative orbit (\Rsunnom) &  \multicolumn{4}{c}{$11.071 \pm 0.040$}         \\
Mass (\Msunnom)                             &   2.087  & 0.022       &   1.972  & 0.024       \\
Radius (\Rsunnom)                           &   1.841  & 0.012       &   1.738  & 0.014       \\
Surface gravity ($\log$[cgs])               &   4.227  & 0.005       &   4.253  & 0.007       \\
Density ($\!\!$\rhosun)                     &   0.334  & 0.006       &   0.376  & 0.008       \\
Synchronous rotational velocity ($\!\!$\kms)&  43.98   & 0.29        &  41.52   & 0.32        \\
Effective temperature (K)                   &   9850   & 280         &   9570   & 270         \\
Luminosity $\log(L/\Lsunnom)$               &   1.459  & 0.050       &   1.358  & 0.049       \\
$M_{\rm bol}$ (mag)                         &   1.09   & 0.12        &   1.34   & 0.12        \\
Interstellar reddening \EBV\ (mag)          &  \multicolumn{4}{c}{$0.103 \pm 0.019$}          \\[3pt]
\end{tabular}
\end{table}


 

The masses, radii and temperatures of the components of \targ\ were compared to theoretical predictions, in the form of the {\sc parsec} 1.2 evolutionary models \cite{Bressan+12mn}. The masses and radii of the stars are well fitted by a solar chemical composition (a fractional metal abundance by mass of $Z=0.017$) and an age of $320 \pm 20$~Myr, but this predicts \Teff\ values too low by 2$\sigma$ for both stars. Instead, a metal abundance of $Z=0.010 \pm 0.002$ and an age of $400 \pm 20$~Myr gives predicted masses, radii and \Teff\ values that match the measured properties of \targ\ to within all error bars. This finding of a slightly subsolar metallicity could be checked using high-resolution spectra.

\section*{Summary and conclusions}

\targ\ is a DEB containing two A0~V stars in a circular orbit of period 2.118~d. We have analysed the light-curve from TESS sector 61 and the RVs published by L82 to determine the masses and radii of the component stars. Our results agree with those from L82 but are more precise due to the much higher quality of the TESS light-curve compared to the ground-based photometry from the 1970s. Spectroscopic orbits from \gaia\ DR3 give masses which are more uncertain and are larger by 2.6$\sigma$; improved results will be obtainable from the forthcoming \gaia\ Data Release 4.

We have introduced a new method to determine the \Teff\ values and interstellar extinction of a DEB, by fitting the SED of the system constrained by the known distance (from \gaia\ DR3), radii and surface gravities (our own measurements) and calibrations of surface brightness versus \Teff. The surface brightness calibrations help by pinning down $A_V$, thus partially solving the degeneracy between interstellar reddening and \Teff.

The eclipse times from the five available TESS sectors show a clear quadratic trend, suggesting the presence of a faint third star on a wider orbit. The available ground-based eclipse times are too scattered to deduce anything further, but forthcoming observations from TESS will help delineate this trend.

Both components of \targ\ are expected to be too hot and luminous to show $\delta$~Scuti pulsations \cite{Murphy+19mn}. We checked this by calculating an amplitude spectrum of the residuals of the {\sc jktebop} fit using the {\sc period04} code \cite{LenzBreger05coast}. No pulsations were found to a 3$\sigma$ upper limit of 0.4~mmag, confirming our expectations.

\targ\ is in the southern observing field (LOPS2 \cite{Nascimbeni+25aa}) of the forthcoming PLATO satellite \cite{Rauer+25exa}, and has been selected as a target as part of the scvPIC process \cite{Zwintz+26exa}. As  result, it will be observed almost continuously for a minimum of two years beginning in mid-2027.


\section*{Acknowledgements}
 
We thank the anonymous referee for a prompt and positive report.
We thank Jerzy Kreiner for providing a list of times of minimum for \targ.
We acknowledge support from STFC under grant number ST/Y002563/1.

This paper includes data collected with the TESS mission, obtained from the MAST data archive at the Space Telescope Science Institute (STScI). Funding for US Institutions for the TESS mission is provided by the NASA Explorer Program. STScI is operated by the Association of Universities for Research in Astronomy, Inc., under NASA contract NAS 5–26555.


This work has made use of data from the European Space Agency (ESA) mission {\it Gaia}\footnote{\texttt{https://www.cosmos.esa.int/gaia}}, processed by the {\it Gaia} Data Processing and Analysis Consortium (DPAC\footnote{\texttt{https://www.cosmos.esa.int/web/gaia/dpac/consortium}}). Funding for the DPAC has been provided by national institutions, in particular the institutions participating in the {\it Gaia} Multilateral Agreement.

The following resources were used in the course of this work: the NASA Astrophysics Data System; the SIMBAD database operated at CDS, Strasbourg, France; and the ar$\chi$iv scientific paper preprint service.

Microsoft Copilot was used as a search tool during this work, but was not used to generate or modify any text in this paper.

For the purposes of open access, the authors have applied a Creative Commons Attribution (CC-BY) licence to any Accepted Author Manuscript version arising from this submission.



\end{document}